\documentclass{mem}
\usepackage{natbib}\usepackage{txfonts}\usepackage{balance}
\usepackage{graphicx}
\usepackage[a4paper]{hyperref}
\idline{00}{000}
\begin{document}

\title{Estimation of X-ray scattering impact in imaging degradation for the SIMBOL-X telescope}

\author{
D. \,Spiga, 
G.\, Pareschi,
R.\, Canestrari,
\and V.\, Cotroneo
     }

\institute{
Istituto Nazionale di Astrofisica --
Osservatorio Astronomico di Brera, Via E. Bianchi 46, I-23807 Merate (LC), Italy --
\email{daniele.spiga@brera.inaf.it}
}

\authorrunning{D. Spiga et al.}

\titlerunning{Estimation of X-ray scattering impact\ldots}

\abstract{The imaging performance of X-ray optics (expressed in terms of HEW, Half-Energy-Width) can be severely affected by X-ray scattering caused by the surface roughness of the mirrors. The impact of X-ray scattering has an increasing relevance for increasing photon energy, and can be the dominant problem in a hard X-ray telescope like SIMBOL-X. In this work we show how, by means of a novel formalism, we can derive a surface roughness tolerance -- in terms of its power spectrum -- from a specific HEW requirement for the SIMBOL-X optical module.

\keywords{X-rays: general -- Telescopes: high angular resolution}}
\maketitle{}

\section{Introduction}
The SIMBOL-X X-ray telescope (\cite{Pareschi06}) will be the first space observatory in the soft ($E <$~10~keV) and hard ($E >$~10~keV) X-ray band with imaging capabilities comparable to the presently operated soft X-ray telescopes Newton-XMM (15~arcsec HEW) and Swift-XRT (20~arcsec HEW). The SIMBOL-X angular resolution goal is $\sim$~15~arcsec HEW at 1~keV and $<$ 20~arcsec HEW at 30~keV.

In order to ensure good imaging capabilities to a hard X-ray grazing-incidence optic like those of SIMBOL-X, some critical points have to be carefully evaluated and controlled:
\begin{enumerate}
\item{{\it Deformation of mirrors profiles, and mirrors misalignment.} The effect on the HEW can be considered as independent of the photon energy $E$, and may be estimated using the geometrical optics (ray-tracing). The impact of this factor is strongly affected by the manufacturing, handling, positioning process of the mirror shells.}
\item{{\it Microroughness of reflecting surfaces.} It causes the scattering of reflected X-rays (XRS) (see e.g. \cite{Church86}). This effect falls in the physical optics domain: it has an increasing relevance for increasing photon energy (decreasing photon wavelength $\lambda$).}
\end{enumerate}
As the SIMBOL-X energy band is extended up to 80~keV and beyond, XRS can be the dominant problem at high energies. Hence, a detailed study has to be carried out to establish microroughness tolerances from the angular resolution mission requirements. In the following sections we show how this can be done.

\section{A relation between the roughness PSD and the HEW($\lambda$) function}

A meaningful surface smoothness requirement depends on the formulation of a reasonable HEW requirement, as a function of the photon wavelength $\lambda$, and should be given in terms of the power spectrum of the microroughness (PSD, {\it Power Spectral Density}, see e.g. \cite{Stover95}). We assume here the final $HEW(\lambda)$ to result from the quadratic sum of two independent contributions: the energy-independent $HEW_{{\mathrm fig}}$ due to figure errors and the HEW X-ray scattering term $H(\lambda)$,
\begin{equation}
	HEW^2(\lambda)\approx HEW_{{\mathrm fig}}^2+ H^2(\lambda).
\label{eq:0}
\end{equation}
We shall hereafter make use of a novel formalism aimed at the direct translation of a given $H(\lambda)$ function into a PSD (\cite{Spiga07}). For a double reflection mirror shell with incidence angle $\theta_{\mathrm i}$, the PSD $P$ at the spatial frequency $f_{\mathrm 0}$ can be computed from the $H(\lambda)$ function along with the equation:
\begin{equation}
	\frac{P(f_{\mathrm 0})}{\lambda}\,\frac{{\mbox d}}{{\mbox d}\lambda}\!\left(\frac{H(\lambda)}{\lambda}\right) = -\frac{\ln(4/3)}{4\pi^2\sin^3\theta_{\mathrm i}}.
\label{eq:1}
\end{equation}
Vice versa, from the numerical integration of a given PSD $P(f)$ we can derive the foreseen $H(\lambda)$ term, always for a double reflection mirror shell:
\begin{equation}
	\int^{\frac{2}{\lambda}}_{f_{\mathrm 0}}P(f)\,{\mbox d}\!f = \frac{\lambda^2\ln(4/3)}{16\pi^2\sin^2\theta_{\mathrm i}},
\label{eq:2}
\end{equation}
in Eqs.~(\ref{eq:1}) and (\ref{eq:2}) the spatial frequency $f_{\mathrm 0}$ is related to the $H(\lambda)$ function as follows:
\begin{equation}
	f_{\mathrm 0}(\lambda)=\frac{H(\lambda)\sin\theta_{\mathrm i}}{2\lambda}.
\label{eq:3}
\end{equation}
The upper integration limit in the Eq.~(\ref{eq:2}) can be modified to account for the finite size of the X-ray detector in the focal plane. To be more conservative, we shall heretofore suppose the detector to be very large to account for photons scattered at large angles, which contribute not only to imaging degradation, but also to effective area reduction.

\section{A possible PSD for SIMBOL-X mirrors}
A realistic microroughness requirement should be derived from a slowly-increasing HEW($E$) function. Therefore, we can initially {\it require} a HEW varying from 15~arcsec at 0.1~keV (essentially mirror profile errors) up to 20~arcsec at 40~keV. From this requirement, we can derive a PSD directly from Eqs.~(\ref{eq:0}) and (\ref{eq:1}), assuming an incidence angle e.g. $\theta_{\mathrm i}$~=~0.18~deg, in the SIMBOL-X mirrors range.

\begin{figure}
\resizebox{\hsize}{!}{\includegraphics[clip=true]{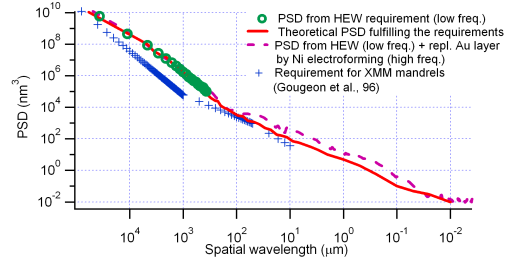}}
\caption{\footnotesize The PSDs under comparison in this work.}
\label{PSD}
\end{figure}

The computed PSD covers the low-frequency regime ($f < (350 {\mathrm \mu m})^{-1}$, the green circles in Fig.~\ref{PSD}). Moreover, from the Eq.~(\ref{eq:2}) at the maximum considered energy (40~keV) we obtain a further constraint on the integral of the PSD at higher frequencies (\cite{Spiga07}),
\begin{equation}
	\sigma=\left[\int^{(0.16 \AA)^{-1}}_{(350 {\mathrm \mu m})^{-1}}P(f)\:\mbox{d}f\right]^{1/2} < 4.2 \,\AA.
\label{eq:4}
\end{equation}
This requirement is quite exacting. For instance, we plot in Fig.~\ref{PSD} the {\it measured} PSD (dashed line) of a Gold layer deposited onto a very smooth master -- an optically polished Zerodur glass -- and replicated by Nickel electroforming. Yet this PSD does not fulfill the Eq.~(\ref{eq:4}), because $\sigma \simeq 6~\:\AA$; hence, it is not suitable to return the required HEW trend. Fortunately, the comparison with the requirements of the XMM mandrels (the "+" signs in Fig.~\ref{PSD}) shows that further improvements should be possible, as better smoothness levels were reached in that case (\cite{Gougeon96}). 

For instance, a {\it simulated} PSD that could fulfill our requirements is represented by the solid line in Fig.~\ref{PSD}. To check this result, we computed from that PSD (Eqs.~(\ref{eq:2}) and (\ref{eq:3})) the expected HEW trend from 1 to 70~keV for three mirror shells of the SIMBOL-X optical module. We assumed 15~arcsec figure error and 3~deg as detector radius, much larger than that of SIMBOL-X ($\sim$~420~arcsec). The resulting HEW values exhibit a moderate increase at low energies, reach a wide "plateau" and diverge at very high energies. The increase is much more marked at the largest incidence angles.

\begin{figure}
\resizebox{\hsize}{!}{\includegraphics[clip=true]{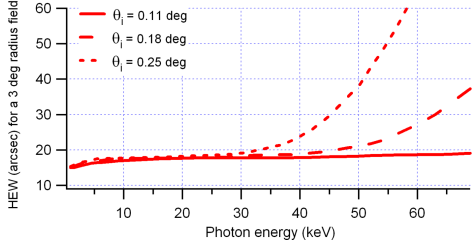}}
\caption{\footnotesize The HEW derived from the PSD model in Fig.~\ref{PSD} (solid line), at three incidence angles of SIMBOL-X mirror shells: 0.11~deg (the smallest one), 0.18~deg (the angle of an intermediate shell, the 55th), and 0.25~deg (the largest one).}
\label{HEW}
\end{figure}

\begin{figure}
\resizebox{\hsize}{!}{\includegraphics[clip=true]{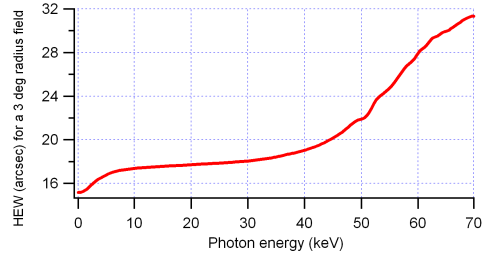}}
\caption{\footnotesize The final HEW for the whole optical module of SIMBOL-X, assuming the solid line in Fig.~\ref{PSD} as PSD of mirrors surface. The HEW is within the requirement.}
\label{finalHEW}
\end{figure}

In order to evaluate the overall angular resolution for the SIMBOL-X optical module, we performed the calculation of the HEW as a function of the photon energy for all mirror shells, assuming the same PSD used in Fig.~\ref{HEW}, and we averaged the HEW values over the respective effective areas $A_k(E)$:
\begin{equation}
	HEW_{\mathrm{tot}}(E)=\frac{\sum_{k=1}^{100} A_k(E)\, HEW_k(E)}{\sum A_k(E)}.
\label{eq:5}
\end{equation}
The final result, plotted in Fig.~\ref{finalHEW}, exhibits a {\it slow increase of the HEW for increasing photon energy from 1 to 70~keV} and fulfills the SIMBOL-X requirement (HEW $<$~20~arcsec at 30~keV). In fact, the damping of the HEW increase is also due to the effective area cutoff at lower energies for larger incidence angles, that balances the steeper divergence of the HEW (see Fig.~\ref{HEW}).

\section{Conclusions}
 The evaluation of the XRS contribution to the HEW in the SIMBOL-X hard X-ray telescope is a very important issue, since at high energies it can be the dominant imaging degradation factor, if the reflecting surfaces are not sufficiently smooth. By means of the formulae mentioned above, a given HEW($E$) requirement -- in the SIMBOL-X energy band -- can be directly translated into important information concerning the surface microroughness tolerance, conveniently expressed in terms of PSD. Conversely, we can use the PSD of the mirror roughness -- measured over a very wide spatial frequency range -- to preview the HEW($E$) function of a single mirror shell, in the range of photon energies and incidence angles of SIMBOL-X. The HEW can be calculated for each mirror shell from the surface PSD, and from that we can easily recover the HEW estimation for the entire optical module. 

 Calculations performed on the PSD of samples replicated from smooth masters by Nickel electroforming proved that the effort has to be concentrated on the damping of the microroughness at spatial wavelengths shorter than $\sim$~300~$\mu$m, in order to improve the angular resolution at high energies and therefore to meet the SIMBOL-X angular resolution requirements in soft and hard X-rays.

\begin{acknowledgements}
O.~Citterio, P.~Conconi, S.~Basso (INAF/OAB) for useful discussions.
\end{acknowledgements}

\bibliographystyle{aa}

\end{document}